\documentclass{aa}  
\usepackage{natbib,twoopt}
\usepackage[breaklinks=true]{hyperref} 
\bibpunct{(}{)}{;}{a}{}{,}             
\makeatletter
  \newcommandtwoopt{\citeads}[3][][]{\href{http://adsabs.harvard.edu/abs/#3}%
    {\def\hyper@linkstart##1##2{}%
     \let\hyper@linkend\@empty\citealp[#1][#2]{#3}}}
  \newcommandtwoopt{\citepads}[3][][]{\href{http://adsabs.harvard.edu/abs/#3}%
    {\def\hyper@linkstart##1##2{}%
     \let\hyper@linkend\@empty\citep[#1][#2]{#3}}}
  \newcommandtwoopt{\citetads}[3][][]{\href{http://adsabs.harvard.edu/abs/#3}%
    {\def\hyper@linkstart##1##2{}%
     \let\hyper@linkend\@empty\citet[#1][#2]{#3}}}
  \newcommandtwoopt{\citeyearads}[3][][]%
    {\href{http://adsabs.harvard.edu/abs/#3}
    {\def\hyper@linkstart##1##2{}%
     \let\hyper@linkend\@empty\citeyear[#1][#2]{#3}}}
\makeatother
\usepackage{graphicx}
\usepackage{txfonts}	
\usepackage{amsmath}	
\usepackage{amssymb}	
\usepackage{enumitem}
\usepackage{multicol}

\usepackage{xcolor}
\usepackage{ulem}

\begin{document}

   \title{A Ly$\alpha$\ nebula at z$\sim$3.3}


   \author{P. Hibon
          \inst{\ref{ESO}},
          F. Tang\inst{\ref{Kapteyn}},
          \and
          R. Thomas\inst{\ref{ESO}}
          }

   \institute{ESO
              Alonso de Cordova 3107, Vitacura, Santiago, Chile \label{ESO}\\
              \email{phibon@eso.org}
              \and
              Kapteyn Astronomical Institute, University of Groningen, Netherlands \label{Kapteyn}\\
             }

   \date{}

 
  \abstract
   {Searching for high-redshift galaxies is a field of intense activity in modern observational cosmology that will continue to grow with future ground-based and sky observatories. Over the last few years, a lot has been learned about the high-z Universe. }
   {Despite extensive Ly$\alpha$\ Blobs (LAB) surveys from low to high redshifts, giant LABs over 100 kpc have been found mostly at z $\sim$ 2-4. This redshift range is coincident with the transition epoch of galactic gas-circulation processes from inflows to outflows at z$\sim$2.5-3. This suggests that the formation of giant LABs may be related to a combination of gas inflows and outflows. Their extreme youth makes them interesting objects in the study of galaxy formation as they provide insight into some of the youngest known highly star forming galaxies, with only modest time investments using ground-based telescopes.}
   {Systematic narrow-band Ly$\alpha$\ nebula surveys are ongoing, but they are limited in their covered redshift range and their comoving volume. This poses a significant problem when searching for such rare sources. To address this problem, we developed a systematic searching tool, ATACAMA (A Tool for seArChing for lArge LyMan Alpha nebulae) designed to find large Ly$\alpha$\ nebulae at any redshift within deep multi-wavelength broad-band imaging. }
   {We identified a Ly$\alpha$\ nebula candidate at $\mathrm{z}_{\mathrm{phot}}\sim$3.3  covering an isophotal area of 29.4$\mathrm{arcsec}^{2}$. Its morphology shows a bright core and a faint core which coincides with the morphology of previously known Ly$\alpha$\ blobs. A first estimation of the Ly$\alpha$ equivalent width and line flux agree with the values from the study led by several groups.}
   {}


   \maketitle
%

\section{Introduction}

The study of galaxy clusters plays an important role in
understanding cosmological structure formation and the
astrophysics of galaxy evolution. Statistics concerning galaxy cluster
size, mass, and redshift distribution provide constraints for
cosmological models, while the properties of the galaxies and
gas inside clusters give clues about galaxy evolution and the
star formation history of the universe \citep[e.g.][]{boylan-kolchin2009}. The progenitors of galaxy
clusters, the so-called protoclusters, start off as over-dense
regions and groups of galaxies at high redshift, which over time
coalesce into the larger galaxy clusters we see today. 
Observing the early stages of
cluster formation at higher redshifts has been challenging.
Since protoclusters lack many of the observational properties
of the massive virialised galaxy clusters of today, one of the
best ways to find them is to identify galaxy over-densities at
high redshift \citep{overzier2016}. Readily observable populations
of galaxies include Lyman break galaxies (LBGs) and Ly$\alpha$\ emitters (LAEs) \citep[e.g.][]{overzier2008}.
These objects, which are compact galaxies that have strong emission in
the Ly$\alpha$ line, are relatively easy to observe over a wide range of
redshifts at z $\sim$ 2–6 \citep[e.g.][]{taniguchi2005,gronwall2007,nilsson2009,guaita2010,Cassata15}.
LAEs are mainly
star-forming, low-mass objects, and some may be the
building blocks of Milky Way-like galaxies \citep{gawiser2007}.\\
Giant Ly$\alpha$-emitting nebulae, also known as Ly$\alpha$ "blobs" (LABs; \cite{steidel2000,matsuda2004}) emit Ly$\alpha$ radiation on large
scales (50--100~kpc) and have high Ly$\alpha$ luminosities of $\sim10^{43-44}$~erg~s$^{-1}$. Moreover, it has been found that they are also apparent occasional signposts of LAE over-densities \citep[e.g.][]{yang2010}.
The mechanism that powers the strong extended Ly$\alpha$ emission in these blobs is
still poorly understood. Possible powering mechanisms include
gravitational cooling radiation \citep[e.g.][]{rosdahl2012}, the resonant scattering of Ly$\alpha$ photons produced by star formation \citep[e.g.][]{cen-zheng2013}, photo-ionising radiation
from active galactic nuclei (AGNs; \cite[e.g.][]{yang2014a}), or a combination thereof \citep[e.g.][]{reuland2003,dey2005}. Another potential
source is shock heating from starburst-driven winds \citep{mori2006}, although recent 
studies of the emission of non-resonant lines from eight
Ly$\alpha$ blobs exclude models that require fast galactic winds
driven by AGNs or supernovae \citep[e.g.][]{prescott2015a}.
Regardless of the energy sources of Ly$\alpha$ blobs, the
association of these blobs with compact LAE over-densities of $\sim$10--20 Mpc in size \citep[e.g.][]{saito2015} suggests that LABs are good potential markers of large
protoclusters. Furthermore, the number density and variance of
Ly$\alpha$ blobs, as well as the 200--400~km~s$^{-1}$ relative velocities of
their embedded galaxies, suggest that blobs themselves occupy
individual group-like halos of  $\sim 10^{13}$~M$_{\odot}$ \citep[e.g.][]{prescott2015b}. 
Thus, blobs may be sites of
massive galaxy formation and trace significant components of
the build-up of protoclusters.\\
These Ly$\alpha$ nebulae are usually found in the redshift range $2<z<6$ using different techniques, such as a narrow-band (NB) search or integral field spectroscopy (IFS).
The narrow-band technique is the most successful method for
detecting strong Ly$\alpha$\ emission lines of galaxies, since it relies
on a specific redshift interval as well as a selected low-sky
background window. Using integral field spectroscopy is an optimal method as it does not require any spectroscopic follow-up and covers not only a large volume but also a wide range of redshifts.\\
To date, a few dozen of blobs have been reported.  \cite{badescu2017} found three new z$\sim$2.3 blobs using Narrow-Band technique. \cite{matsuda2004} and \cite{arrigoni2019} found several tens of these objects around a redshift of z$\sim3$. \cite{vanzella2017} and \cite{lusso2019} found five LABs using the IFS technique with the VLT/Multi-Unit Spectroscopy Explorer (MUSE) around z$\sim$3.3. \cite{borisova2016} detected one at z$\sim$3.5, and \cite{saito2015} also detected one at z$\sim$4.1. The highest redshift LAB remains Himiko at z$\sim$6.6 discovered by \cite{ouchi2009}, also found using the narrow-band technique.\\
This paper is organised as follows : in Section~\ref{sec:method}, we first present the method and the data set we used. In Section~\ref{sec:lab}, we explain the characteristics of the newly detected Ly$\alpha$\ nebula. 
Throughout this study, we adopted the following cosmological parameters: $\mathrm{H}_{0}$ = 70 km s$^{-1}$ Mpc$^{-1}$, $\Omega_{m}$ = 0.3, and $\Omega_{\Lambda}$= 0.7. All magnitudes are given using the AB system \citep{Oke83}.
\section{Method and data set}\label{sec:method}
The ATACAMA project (PI : P.\ Hibon; Hibon et al.\ 2020, in prep.) 
aims to perform a systematic search for Ly$\alpha$\ nebulae in a multi-wavelength broad-band imaging data set, instead of using the conventional narrow-band imaging technique. This strategy is similar to
the one described in \cite{prescott2012} who pioneered this method by confirming the detection of several LABs at $1.7\leq z \leq 2.7$ \citep{prescott2013}.
Based on this method, we designed a software (Thomas et al.\ 2020, in prep.) that consists of a GUI communicating with SExtractor \citep{bertin1996} and optimised for LAB detection. 
We use a single MegaCam ultra deep pointing of 1 sq. degree of the Cosmic Evolution Survey (COSMOS, \citealt{Scoville2007}) field from the CFHT Legacy Deep Survey (CFHTLS Deep), reaching limiting magnitudes between 25 and 26.44 depending on the filter (80\% completeness limit). This provides a high-quality and homogeneous optical data set, calibrated photometrically to 1.0\% and astrometrically to 0.028~arcsec. This data set was reduced by Terapix dedicated to the processing of very large data flows from digital sky surveys. The transmission curves are available online for these \href{https://www.cfht.hawaii.edu/Instruments/Imaging/Megacam/specsinformation.html}{optical} and \href{https://www.cfht.hawaii.edu/Instruments/Filters/cfhtir.html}{near infrared} filters.\\
We selected the COSMOS field due to the availability of a very wide range of wavelength bands in various data archives. \\
The ATACAMA tool processed the images through various steps:
(1) the progressive elimination of bright, intermediate, and faint point-like objects in one specific band; 
(2) the production of wavelet planes allowing us to detect extended objects; 
(3) the  creation of a photometric candidate catalogue with the spectral energy distribution in all available broad-band images; 
(4) the cross-correlation of the catalogue of candidates with the COSMOS2015 photometric redshift catalogue (\cite{laigle2016}), which
contains photometric redshifts for over half a million sources in the CFHTLS Deep2 field;
(5) the determination of the photometric redshift through spectral energy distribution (SED) fitting with LePhare (\cite{arnouts2011}) in case of no matching object in the COSMOS15 catalogue;
(6) the careful visual inspection of each object.
These steps are performed by several team members and the final results are compared between the different parameter choices of each step.

\section{The Ly$\alpha$\ nebula}\label{sec:lab}
One object (LAB1) in particular caught our immediate interest, after we ran ATACAMA several times and independently of the parameter variations during the different tests. 


 
\subsection{Photometric information}

\begin{figure*}[!h]
\includegraphics[width=\textwidth]{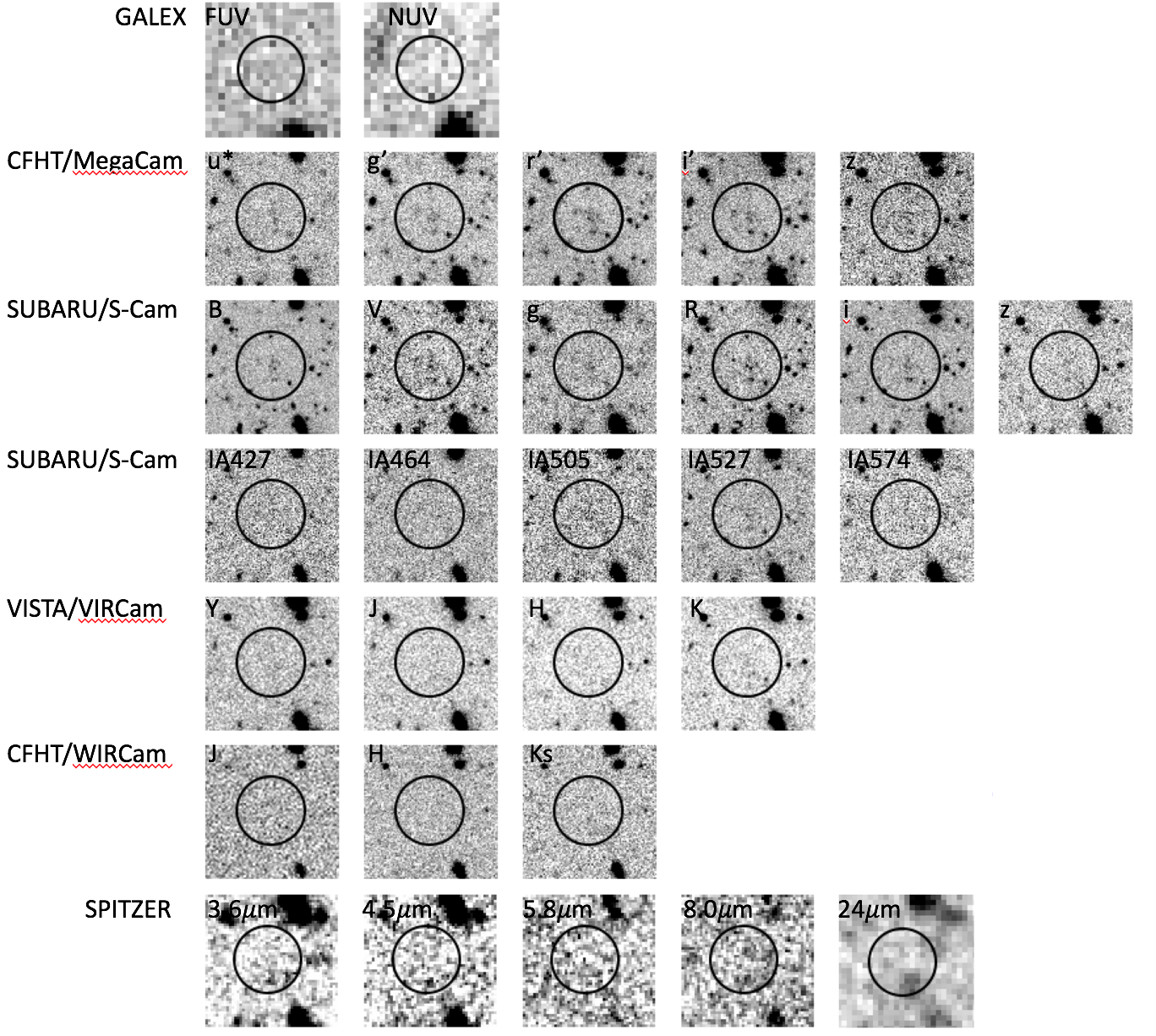}
   \caption{Thumbnails. North up, east left. Size of each stamp is 28 arcsec side-wise. We show here NUV and FUV from GALEX, the optical bands from CFHT/Megacam, the bands available from SUBARU/Suprime-Cam, the NIR bands from VISTA/VIRCam and from CFHT/WIRCam, the four channels from Spitzer/IRAC and the 24$\mu$m Spitzer/MIPS. The circles are 7.65 arcsec in radius and are encompassing the complete nebula.}
      \label{fig:thumbnail}
\end{figure*}

As mentioned in the previous section, SExtractor is used to estimate the photometric information of this object. We chose to use the isophotal-corrected magnitudes as such an extended object does not have all its flux within neat boundaries. The double-image mode is used to derive the multi-band photometry of this LAB, with the g'-band image as the detection image and all other bands as photometric bands. 
Figure~\ref{fig:thumbnail} shows the multi-wavelength $28"\times28"$ images of the LAB. We choose not to show the Herschel/SPIRE stamps at 250microns, 350microns and 500microns, as their resolution is not optimal to possibly find a detection (6"/pixel, 10"/pixel, 14"/pixel, respectively).
\cite{prescott2012} selected their highest  $2\leq z \leq 3$ LAB candidate with the following colour : $B_{W} - R < 0.45$ while searching for LABs in the $B_{W}$ band image. We verified the colour we obtained from its magnitudes. 
This LAB has $g'-r' = 0.45$. The dropout has a magnitude difference of $u'-r' = 2.71$. 
Our object is neither detected in J-, H- nor Ks-band images. 
\cite{vanderburg2010} led a search for Lyman-break Galaxies in the CFHTLS data. They defined the following criteria to detect u-dropout candidates: 
\begin{equation} 
\begin{split}
1.0 & < (u^{\ast}-g')\\
-1.0 & < (g'-r') < 1.2 \\ 
1.5(g'-r')  & < (u^{\ast}-g')-0.75. \nonumber
\end{split}
\end{equation}

\noindent
The colours of our newly detected Ly$\alpha$ nebula respect their criteria and reinforce the detection and the high-redshift nature of this object. 

\begin{table*}[]
    \centering
    \begin{tabular}{|c|ccc|}
    \hline
        Band & Complete Nebula & Bright Core & Faint Halo \\
        RA &  09:59:54.913  & 09:59:54.658 & 09:59:54.9263\\
        Dec & +01:56:12.389 & +01:56:08.81 & +01:56:12.297 \\
    \hline
      CFHTLS-u* & 27.64$\pm$0.81& 26.47$\pm$0.13 & 27.83$\pm$0.41\\
      CFHTLS-g' & 27.68$\pm$1.29& 26.22$\pm$0.08& 27.33$\pm$0.19\\
      CFHTLS-r' & 25.25$\pm$0.22& 26.03$\pm$0.09 & 26.74$\pm$0.15\\
      CFHTLS-i' & 25.3* & 25.94$\pm$0.12 & 27.75$\pm$0.54\\
      CFHTLS-z & 25.1* & 24.73$\pm$0.09 & 27.53$\pm$1.04\\
      Subaru B & 26.26$\pm$0.09 & 26.58$\pm$0.05 & 27.15$\pm$0.08\\
      Subaru V & 25.56$\pm$0.16& 26.28$\pm$0.08 & 26.7$\pm$0.12\\
      Subaru g & 25.90$\pm$0.10& 26.62$\pm$0.05& 26.82$\pm$0.10\\
      Subaru IA527 & 25.31$\pm$0.11 & 26.21$\pm$0.12 & 26.41$\pm$0.12\\
      Subaru R & 25.2$\pm$0.10& 26.14$\pm$0.06 & 26.57$\pm$0.08\\
      Subaru i & 24.86$\pm$0.05& 26.20$\pm$0.09 & 26.33$\pm$0.10\\
      Subaru z & 24.92$\pm$0.16& 25.9* & 26.97$\pm$0.45\\
      WIRDS J & 24.1$\pm$0.47& 23.2*& 21.7*\\
      WIRDS H & 20.9*& 23.5* & 24.95$\pm$0.48\\
      WIRDS Ks & 20.8*& 23.4* & 25.46$\pm$0.94\\
      VISTA Y & 24.46$\pm$0.24& 25.7* & 25.94$\pm$0.11\\
      VISTA J & 24.7* & 25.3* & 25.86$\pm$0.30\\
      VISTA H & 24.43$\pm$0.14 & 25.0* & 26.19$\pm$0.67\\
      VISTA Ks & 23.63$\pm$0.08 & 24.94$\pm$0.11 & 25.48$\pm$0.1\\
      Spitzer 3.5$\mu$m & 22.9* & 25.5* & 23.8*\\
      Spitzer 4.6$\mu$m & 22.9* & 25.5* & 23.8*\\
    \hline
    \end{tabular}
    \caption{Aperture magnitudes of the complete nebula : 35 pixels aperture, 3$\sigma$ detection, the bright core : 3arcsec aperture, 3$\sigma$ detection, and the faint halo : 15 pixels aperture, 3$\sigma$ detection. The magnitude marked with * are limiting magnitude at a 3$\sigma$ level with an error of $\pm$0.1.}
    \label{tab:mag}
\end{table*}


\subsection{Morphology of LAB1}
\begin{figure*}[h!]
\centering
\includegraphics[width=\textwidth]{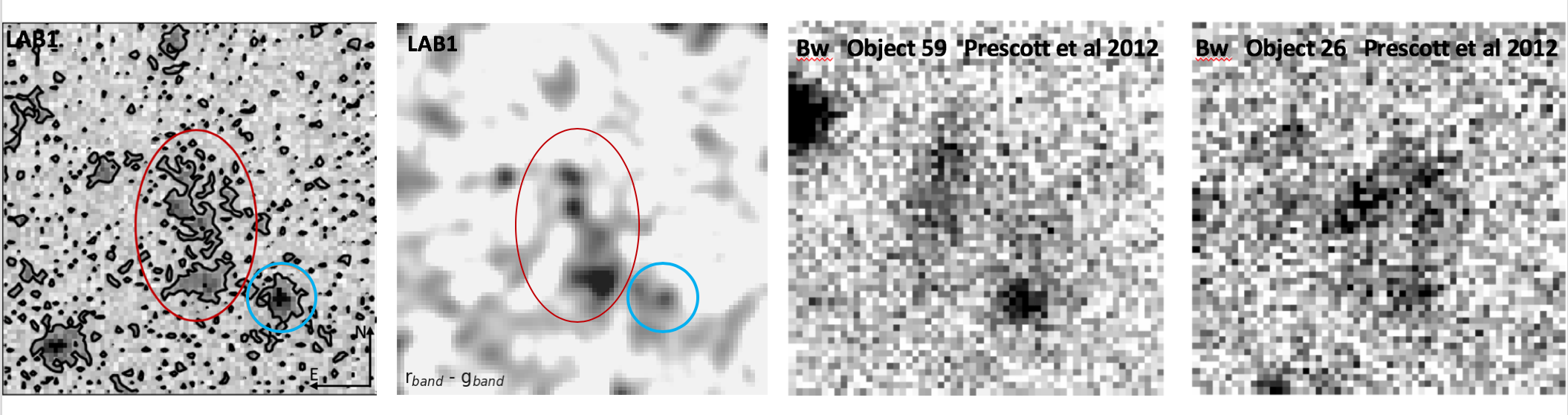}
   \caption{Left : contour of LAB1 on the 15"$\times$15" r-band image. North is up, east is left. The identified bright core is encircled in blue. The filamentary structure is encircled in red. The second image shows the r-g bands subtraction. The third image is the Bw-band of object 59 of \cite{prescott2012}, Right : Bw-band of object 26 of \cite{prescott2012}. Both Bw-band stamps were extracted from Figure 10 of \cite{prescott2012}}
      \label{fig:contour}
\end{figure*}

This LAB is composed of one main clump and a halo, as can be seen in the g'- and r'-band (see Fig. \ref{fig:thumbnail}). Its complete extent is 28 kpc x 59 kpc, which corresponds to $29.4 \mathrm{ arcsec}^{2}$, measured in the g'-band image.  
The main clump, also called the bright core, is located at RA=149.97769, Dec=1.9358. The halo or filamentary structure is orientated 45deg north-east of the bright core, located at RA=149.9777428, Dec=1.9357793, as seen and indicated in Figure~\ref{fig:contour}, where we show the contour of this LAB with the filamentary structure encircled in red ( area of the red ellipse : 24.3 sq.arcsec) and the bright core in blue (area of the blue circle : 5.4 sq.arcsec). We also show in the same figure the r'-band image subtracted by the g'-band. \\ 
The morphology of this Ly$\alpha$\ nebula is very similar to the famous one found by \citet[Fig.~6]{steidel2000} (called Blob 1), as it can be resolved into a bright source and a diffuse gas halo. The bright core is not centred on the peak of the extended emission but it is located at 5.58 arcsec in the south-western direction (see Fig~\ref{fig:contour}). 
There is no established consensus on the relation morphology-powering source yet. Both radio-loud and radio-quiet LABs can show filamentary and/or bubble-like structures \citep{matsuda2004,dijkstra2009,geach2009}. 
\cite{prescott2012} divided their candidates in two morphological categories: GROUP and DIFFUSE.
GROUP corresponds to a tight grouping of compact sources selected as a single large source, and DIFFUSE indicates spatially extended diffuse emission. When comparing our object (Figures~\ref{fig:thumbnail} and~\ref{fig:contour}) to the complete sample of \cite{prescott2012}, we first remark that it is more similar in morphology with the DIFFUSE ones. 
From Figure 10 of \cite{prescott2012}, 21 out of 85 blobs are diffuse.
Comparing this sub-sample with LAB1 in more detail, it presents the same kind of structure as source numbers : 5, 10, 26, 59. Figure~\ref{fig:contour} shows the detection band for objects 26 and 59 on the side of our r'-band contoured object, to facilitate the visual morphology comparison. \\
The morphologies in the sample of \cite{matsuda2011} are very diverse, going from compact systems to filamentary, bubble-like, conical structures, including some with shapes similar to our candidate.\\
\cite{borisova2016} and \cite{arrigoni2019} Ly$\alpha$\ nebulae with a very circular morphology, centred on the host quasi-stellar objects (QSOs). 
However, blob 2 of \cite{borisova2016} is elongated, with the QSO on one of the extremity. It could also be seen as a more conical structure. From both samples, this is another very similar object to LAB1. \\
\cite{cai2017} presented a blob at z=2.3 with a filamentary structure possibly aligned with a large-scale structure found previously \citep{Cai2016}.\\
The newly discovered z=3.32 LAB from \cite{marques2019} shows an elongated morphology (and is extended over $\simeq$11", or $\simeq$85 kpc at z = 3.33, within the 3$\sigma$ detection limit.)\\
\cite{reuland2003} reported the detection of Ly$\alpha$\ nebulae centred on z$>$3 radio galaxies, two at z$\sim$3.8, one at z$\sim$3.4. We focus our comparison on B2 0902+34 (z$\sim$3.4) which shows a diffuse morphology with an extension of 10"$\times$8" and some possible signs of large-scale filamentary emission (Figures 7 and 8 of \cite{reuland2003}).\\

\cite{vanzella2017} reported a z$\sim$3.3 LAB in the Hubble Ultra Deep Field (HUDF) with a size of $\sim$ 40 kpc $\times$ 80 kpc. \cite{lusso2019} found three blobs at z$\sim$3.2 : the most extended one has a 150kpc and the smallest one 35 kpc. The measured size of our newly found object is therefore in agreement with the Ly$\alpha$ nebulae previously found in a similar redshift range.

\subsection{Ancillary data}
Based on the position of our blob candidate we looked for counterparts from different astronomical resources. We list below the result of this search from different facilities.\\

\noindent
\textbf{HST} : the Cosmic Assembly Near-infrared Deep Extragalactic Legacy
Survey (CANDELS) data set \citep{grogin2011,koekemoer2011} covers 210 sq.arcmin in the following ACS (Advanced Camera for Surveys) and WFC3 (Wide Field Camera 3) filters : F606W, F814, F350LP, F125W and F160W. However, the CANDELS area does not overlap with our object. \\

\noindent
\textbf{CFHT} : the CFHTLS also provides the WIRCam Deep Survey (WIRDS, \cite{bielby2012}), an extremely deep, high-quality (FWHM$\sim$0.6") J, H, Ks imaging covering a total effective area of 2.1 sq.deg and reaching AB 50\% completeness limits of $\sim$24.5.\\

\noindent
\textbf{CHANDRA} : we verified the available merged images \citep{Elvis2009} corresponding to the X-ray energy bands 0.5 - 2 keV (soft), 2 keV - 7 keV (hard), and 0.5 - 7 keV (full) energy bands, reaching $2.2. 10^{-16}$ cgs, $1.5. 10^{-15}$ cgs and $8.9. 10^{-16}$ cgs in depth, respectively. There is no detection associated with our object.\\

\noindent
\textbf{GALEX} : we also searched for a possible counterpart in the near and far-UV images reaching a depth of 26 AB magnitude \citep{Zamojski2007}. 
We did not identify any detection in these bands.\\

\noindent
\textbf{SUBARU} : we checked the SUBARU/Suprime-Cam mosaic images available in several filters \citep{Taniguchi2015}. We detected our object in the Subaru mosaic with the intermediate filter IA527, centred at $5262.3\AA$, FWHM=243.5\AA.  As the photometric redshift of our Ly$\alpha$ nebula is 3.32 (see Sect. \ref{photoz}), we should expect to have a detection of the extended emission at $\lambda \sim 5225\AA$, wavelength covered by this intermediate band filter. The Ly$\alpha$ nebula is detected in this image.\\ 

\noindent
\textbf{SPITZER} : looking at a possible detection in IRAC channels 3.6 and 4.5 microns, and in the MIPS image at 24 microns, we found a marginal detection in the 24$\mu\mathrm{m}$ MIPS image \citep{Sanders2007} at RA=149.97831, Dec=1.8353, at 2.86arcsec south-east of the bright core of our nebula. Hence, we can assume that this could be a marginal detection of the bright core of this LAB. The 24$\mu\mathrm{m}$ MIPS image has a 3$\sigma$ detection limit of 54$\mu$Jy. Using the zero-point information given in the MIPS Instrument Handbook, we found a flux of 29.9$\mu$Jy$\pm$5.9 for the possible bright core detection. \\

\noindent
\textbf{HERSCHEL} : we looked for a possible detection in the SPIRE images reaching the following 5$\sigma$ noise levels: 8.0mJy, 6.6mJy, 9.5mJy in the SPIRE bands 250, 350 and 500 mJy, respectively \citep{Oliver2012,Chapin2011,Levenson2010,Viero2013,Viero2014}.  The very low resolution of the SPIRE images (6"/pixel,10"/pixel and 14"/pixel for the 250, 350 and 500 mJy bands respectively) prevent us from concluding on the possible detection for any counterpart corresponding to the associated bright core, which could be the host galaxy. \\
We thus cannot rule out the presence of an associated active galactic nucleus, which could be the powering source of this extended Ly$\alpha$ emission.\\

\noindent
\textbf{VLA} : we verified the available VLA images corresponding to the following frequencies : 327 MHz (P-band, 90cm), 1.4 GHz ( = L-band, 20cm) \citep{Schinnerer2007,Bondi2008}, and S-band at 2-4 GHz \citep{Smolcic2017}. These images reach the following 3$\sigma$ detection limit : 0.12mJy in P-band, 21$\mu$Jy in L-band and 3.75mJy in S-band. There is no VLA detection associated either with the bright core or the filamentary component of our object.\\

\noindent
\textbf{ALMA} : we looked for corresponding radio sources in the ALMA Science Archive. We did not find any ALMA data taken in the surroundings of our LAB.

\subsection{Photometric redshift}
\label{photoz}
\begin{figure}
\centering
    \includegraphics[width=9cm]{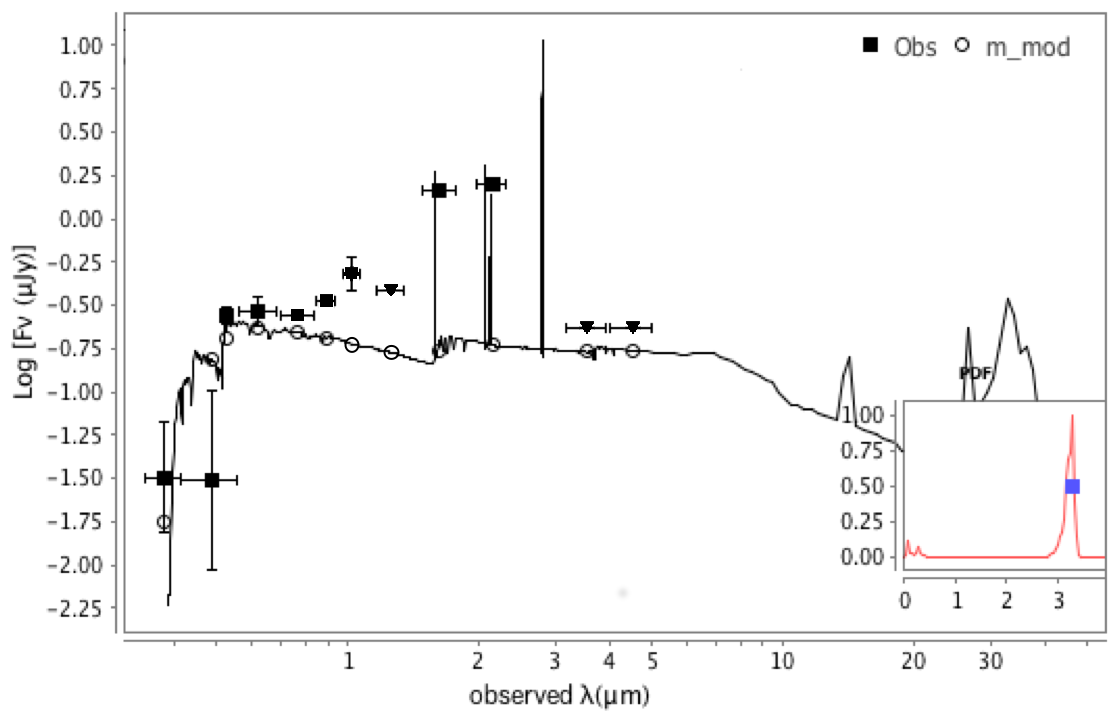}\par 
    \includegraphics[width=9cm]{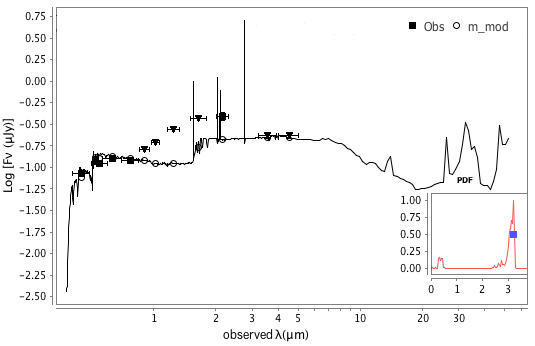}\par 
    \includegraphics[width=9cm]{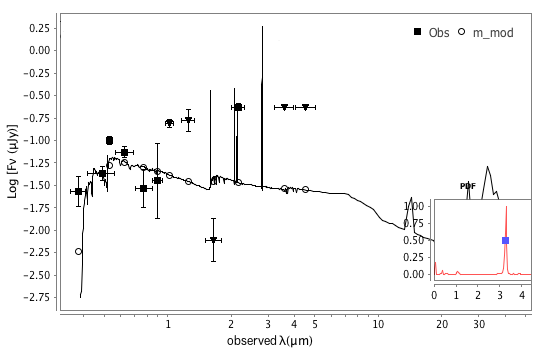}\par 
\caption{Best-fit galaxy template in observed wavelengths from LePhare including the emission line contribution. Twelve filters were used to estimate these photometric redshift. The filled squares represent the ISOCOR magnitudes of our object, and the empty circles show the magnitudes from the best model. The downward facing triangles represent the detection limit used for the photometric redshift estimation. Top : for the complete nebula, magnitude of aperture 35 pixels. Centre : for the bright core, with 3arcsec aperture, 3sigma detection. Bottom : for the faint halo with 15 pixels aperture. The three fits give a redshift of around 3.3$\pm$0.05.}
\label{fig:lephare2}
\end{figure}

The COSMOS2015 catalogue \citep{laigle2016} contains precise photometric redshift for more than half a million objects over the whole COSMOS field. They derived them using the photometry of 30 bands, several tens of templates and several extinction laws. For a redshift in the range $3<z<6$, they reached a photometric redshift precision of 0.021.
We verified any possible corresponding object for this LAB, and found no matching entry in the COSMOS2015 catalogue within a search radius of 2 arcsec.
It is also absent from the COSMOS Master Spectroscopic Redshift Catalogue, which contains more than 65000 sources (M. Salvato, private conversation). Hence, we need to estimate a photometric redshift.\\
We fitted its SED using the PHotometric Analysis for Redshift Estimations, also known as LePhare, \citep{arnouts1999}. The photometric redshift was then computed using LePhare \citep{arnouts2011}, which is based on a $\chi^{2}$
template-fitting method \citep{arnouts2011}. We used the galaxy templates optimised for the COSMOS survey \citep{ilbert2006} with the emission line contribution enabled. \\
Table~\ref{tab:mag} presents the $3\sigma$ magnitudes for the complete nebula, the bright core and the faint halo in all the available filters.
We used the following bands to obtain necessary photometric information for the bright core, the faint core, and the complete nebula. 
We used aperture magnitudes obtained with SExtractor with a diameter aperture of 3 arcsec for the bright core, 15 pixels for the faint part, and 35 pixels for the whole nebula. \\
A $3\sigma$ detection threshold is used to determine the magnitudes for all of them.\\
For the bright core, different tests of filter combinations were realised with a minimum of 11 filters used each time. For some of these tests, we used the flux measured in the $24\mu\mathrm{m}$ MIPS image, but we chose not to include it in the figure as we have no certainty it really corresponds to the bright core. All the tests are consistent with a result of $\textrm{z}_{\textrm{phot}}=3.32\pm0.05$ with $\chi^{2}$=25.9.
Similarly to the bright core, we also realised several tests with different apertures (10-, 15- and 20-pixel diameters) and different filter combinations to estimate the photometric redshift of the faint core, taking advantage of the suit of data available in the COSMOS field and of the marginal MIPS24$\mu$m detection. All the tests are consistent with a resdhift estimation of $\textrm{z}_{\textrm{phot}}=3.28\pm0.03$  with $\chi^{2}$=20.88.
The photometric redshift of the complete nebula was calculated with an aperture of 35 pixels in diameter. The same strategy was employed by using different combinations of filters. All confirmed a photometric redshift of  $\textrm{z}_{\textrm{phot}}=3.3\pm0.05$  with $\chi^{2}$=15.67. The resulting fits are shown in Figure~\ref{fig:lephare2}.\\

\subsection{Ly$\alpha$\ line}
With the applied broad-band selection technique, estimating the 
Ly$\alpha$ equivalent width is a delicate task. Moreover, we know that all the flux coming from the g-band cannot be fully accounted to the Ly$\alpha$\ line.\\
Based on the redshift estimation of the previous section, the Ly$\alpha$\ line, at $\sim$5248\AA~should be present in the intermediate SUBARU filter IA527. Assuming that all the flux in this filter is associated with the Ly$\alpha$\ emission, we compute that the Ly$\alpha$\ flux is $\sim1.5\times10^{-15}$ erg.s$^{-1}$.cm$^{-1}$. Obviously, this should be considered as an upper limit. We also tried to estimate the Ly$\alpha$ equivalent width using broad-band SED fitting.
For this exercise, we used the SPARTAN software\footnote{https://astrom-tom.github.io/SPARTAN/build/html/index.html} (Thomas, A\&C, submitted). SPARTAN is an SED fitting algorithm based on a $\chi^2$ statistic. It is able to use both photometry and spectroscopy, either separately or at the same time to infer the physical parameters of galaxies.
For the purpose of the Ly$\alpha$ equivalent width estimation, we used \citet{BC03} models at low resolution with the \cite{Chab03} initial mass function. The metallicity can take values up to the solar metallicity (Z$_{\odot}$). We used a star formation history exponentially delayed with a timescale parameter from 0.1 to 2.0 Gyr. In terms of attenuation, we used the Calzetti's dust attenuation \citep{Calzetti00} with E(B-V) ranging from 0.0 to 0.5, and we used the IGM prescription of \citet{Thomas17b} which made it possible to treat the IGM as a free parameter.
SPARTAN includes the possibility to add emission lines. This is done following the method proposed in the paper by \citet{Daniel09}. In this method both nebular continuum and emission lines are added converting the number of ionising photons to H$\beta$ flux. Then, the other emission lines are added using the metallicity dependent line ratios given by \citet{AndersFritz03}. While these line ratios are fixed during the fitting process, it is possible to test multiple possibilities by changing them manually. This allows us to compute the equivalent width directly on the theoretical model for different line ratios. We made the Ly$\alpha$/H$\beta$ ratio vary between 2 and 40 in steps of 2. At each step, we computed the equivalent width in the theoretical model and estimated the $\chi^2$. \\
\begin{figure}[h!]
\centering
\includegraphics[width=\columnwidth]{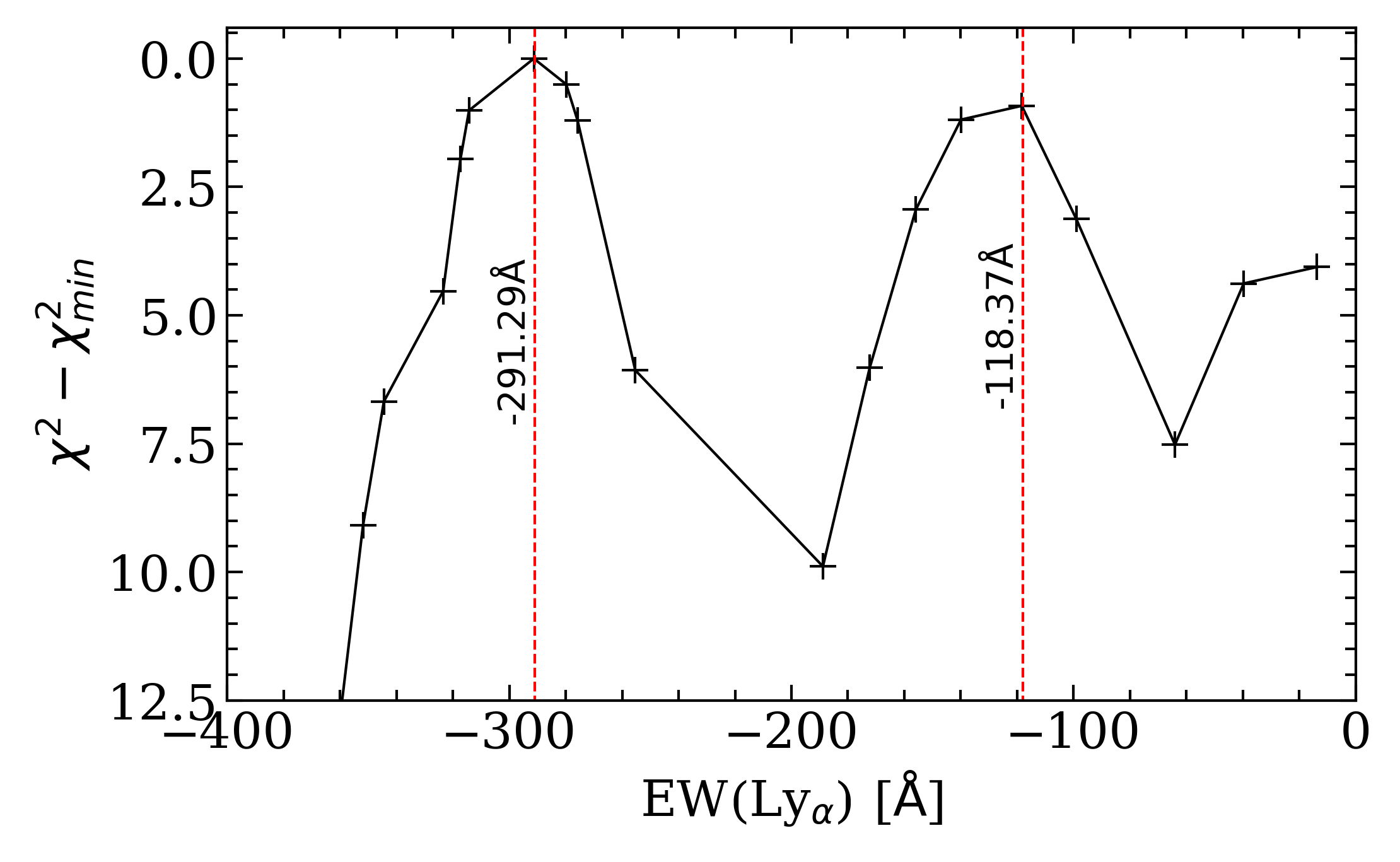}
   \caption{Attempt to estimate the equivalent width of Ly$\alpha$ based on the variation of the line ratios during SED fitting. The two vertical dashed lines give the position of the two local minima. The y-axis has been reversed for clarity and we choose the emission line to be negative.}
      \label{fig:EW}
\end{figure}
The results are displayed in Fig.\ref{fig:EW} where we display the evolution of the $\chi^2$ from the SPARTAN fit with respect to the equivalent width of Ly$\alpha$ (computed in the observed frame). It shows that there are two $\chi^2$ minima at EW(Ly$\alpha$)$\sim$-291\AA~and EW(Ly$\alpha$)$\sim$-118\AA~. This corresponds to line fluxes of 2.02$\times10^{-15}$ erg.s$^{-1}$.cm$^{-1}$ and 1.38$\times10^{-15}$ erg.s$^{-1}$.cm$^{-1}$, respectively.  The difference in $\chi^2$ for these two solutions is less than 1. This is in good agreement with the estimation from the SUBARU band presented above. In addition, these values are consistent with the LAB search of \citet{matsuda2004} whose few candidates have a line flux $\sim10^{-15}$ erg.s$^{-1}$.cm$^{-1}$. The two nebulae, identified by \cite{steidel2000} also have a typical Ly$\alpha$\ flux of the order of $\sim1.3\times10^{-15}$ erg.s$^{-1}$.cm$^{-1}$ (for a physical extent of $\sim$100kpc). \cite{lusso2019} estimated the Ly$\alpha$\ flux  for >100 kpc blobs and found them to be at $\sim0.2-0.8\times10^{-15}$ erg.s$^{-1}$.cm$^{-1}$. Finally, \cite{marques2019} found a Ly$\alpha$ flux of of $\sim1.8\times10^{-15}$ erg.s$^{-1}$.cm$^{-1}$ for the giant blob they found that extend to $\sim$100kpc. Our estimations will have to be verified using spectroscopic data.




\section{Environment and contaminants}

In this section, we discuss the different aspects of the detection of our candidate.\\

\noindent
\textbf{Possible contaminants}: there are several populations that may contaminate our Ly$\alpha$ nebula candidate sample. The morphological selection method tends to select any source of low surface-brightness extended emission that cannot be easily identified, or low surface-brightness nebulae within the Galaxy. In principle, lower redshift sources such as $[OII]\lambda\lambda$3727,3729 emitters,  are a contaminant population in Ly$\alpha$-emitting galaxy surveys \citep[e.g.][]{hibon2010,hibon2011,hibon2012}.
The $[OII]$ emitters would have a redshift in the range [0.065;0.55] considering the transmission of the CFHT/MegaCam g'-band filter with a mean redshift of z$\sim$0.3 for $\lambda_{\mathrm{mean}}\sim4862.5\AA$. The lack of detection in the broad-band COSMOS HST data very likely rules out all of the low-redshift options. Thanks to the result from LePhare, we can hence discard these usual contaminants in such a broad-band search. \\

\noindent
\textbf{Quasar}: 
 using quasar templates during the photometric redshift estimation, we find a photometric redshift of z $\sim 3.3\pm0.05$. This reinforces the assumption that this Ly$\alpha$ nebula could be associated with a quasar at a similar redshift, as explained in \cite{borisova2016} and \cite{Husemann2018} among others. \cite{borisova2016} searched for giant Ly$\alpha$ nebulae around 12 of the brightest radio-quiet quasars. \cite{Husemann2018} studied the environment and the extended Ly$\alpha$ nebula of one specific luminous radio-quiet QSO.  
For such a redshift of z$\sim$3.3, we also expect to have the Lyman limit at $\lambda\ \sim 3921$\AA, which agrees with our SED (see Fig.~\ref{fig:lephare2}).\\

\noindent
\textbf{Radio Detection:} the absence of any radio detection informs us that there is no radio core associated with the nebula. The marginal detection of the bright core in the MIPS 24$\mu$m image with the absence of any radio detection could indicate the presence of a radio-quiet quasar associated with the bright core. However, with its very limited resolution, the MIPS image does not provide more information.\\
\cite{Sargent2010}  differentiate star-forming galaxies and AGNs from their (u-K) colour : AGN sources should show (u-K) $\geq$ 2.42. The bright core of our LAB has a u-K=1.53. Hence, following their criterion, it is likely that the bright core of LAB1 is a star-forming galaxy rather than an AGN. Using their definition of $q_{24,obs}$ (Equation 2 of \cite{Sargent2010}), determined as the  ratio of observed  MIPS 24$\mu$m and VLA 1.4 GHz fluxes, and their Figure 11 showing $q_{24,obs}$ in function of redshift for star-forming and AGN, we remark that at z$\sim$3.3, star-forming galaxies would have $q_{24,obs} < 1$ and AGN $q_{24,obs} \sim 1$. Using the VLA detection limit for our bright core, we have an upper limit of its FIR/Radio ratio of $q_{24,obs} \sim 1.42$. Data with a better FIR resolution and deeper 1.4GHz would help us to establish better constraints on the nature of LAB1's bright core.\\

\noindent
\textbf{Clustering:} from the COSMOS spectroscopic catalogue (private communication with M. Salvato), there is no evidence of a cluster at a similar redshift or at another redshift in the vicinity of our object within a one arcmin radius. The redshift distribution in this same area is in the range of z=0 to 1.5. No source is included in the COSMOS spectroscopic catalogue within 30 arcsec radius from the nebula, except three foreground objects all at z<1 : (09:59:545461; +01:56:36.395), (09:59:54.4717; +01:56:22.861) and (09:59:54.4967; +01:55:59.426).
From the COSMOS2015 photometric redshift catalogue \citep{laigle2016}, in a one arcmin radius around this LAB, there are 61 galaxies with an estimated photometric redshift in the range of 0.05$<\mathrm{z}_{\mathrm{phot}}<$3.68. Eight of these objects have $\mathrm{z}_{\mathrm{phot}}\geq3$. Five of them are in a 2.4 arcmin radius of each other, at 36 arcsec from our object. These five objects have a photometric redshift in the range $3.035< \mathrm{z}_{\mathrm{phot}}<3.68$. \\
From \cite{Ono2018}, the only spectroscopically confirmed Ly$\alpha$ emitter at z$<$4, and more precisely at z=3.41 is 8.41 arcmin north-east of our object. We also verified the presence of  Ly$\alpha$ emitter at z$\sim$3.23 selected in the Subaru/IA527 filter and presented in \cite{Sobral2018} and \cite{Santos2020}. They created and studied a catalogue of 641 z$\sim$3.23 LAEs. Their four closest z$\sim$3.23 LAEs are at the following separations : 1.79arcmin S-E, 1.82 arcmin N-W, 1.96 arcmin N-E and 2.56arcmin S-W.\\
The nebula discovered by \cite{marques-chaves2019} and the B2 0902+34 nebula detected by \cite{reuland2003} do not have a possible cluster or protocluster in their proximity. As mentioned previously, \cite{borisova2016} searched especifically around bright QSOs, \cite{matsuda2011} in the SSA22 protocluster, and \cite{cai2017} in a Ly$\alpha$ absorption system. Their search and selection focused on LABs detected in specific kinds of surroundings. 
The absence of a cluster at similar redshift in the vicinity of our nebula candidate does not affect its legitimate detection. Its spectroscopic confirmation is essential to assessing the complete situation.

\section{Conclusions}

This paper presents the first result obtained from the ATACAMA project (PI: Hibon).
   \begin{enumerate}
      \item We ran a broad-band search technique dedicated to searching for Ly$\alpha$\ nebulae, presented in \cite{prescott2012}. After cross-correlating our candidates with the COSMOS2015 photometric redshift catalogue, we used the LePhare tool to determine the photometric redshift of the candidates absent from the COSMOS2015.  
      \item We identified in the g'-band filter, from the CFHTLS COSMOS data set, one particular Ly$\alpha$\ nebula covering an isophotal area of 29.4$\mathrm{arcsec}^{2}$. Morphologically, we can identify a bright core and a faint core. Its redshift was photometrically estimated at z$\sim$3.3. We obtained a first estimation of the Ly$\alpha$ equivalent width and line flux. Both agree with the values from the Ly$\alpha$ nebulae found and spectroscopically observed by \cite{matsuda2004}.  
      \item With VLT/XSHOOTER spectroscopy we will confirm the redshift of this exciting LAB, which will then be part of the high redshift Ly$\alpha$\ blob sample discovered in a blind search. We are planning to use the following slit widths : 1.3", 1.2" and 1.2" for the UVB, VIS, and NIR arms, respectively. A position angle of 100deg will be ideal to observe both the bright core and part of the faint halo. We will set up the exposure times at 19$\times$600 seconds for each arm.\\ The study of such an object can significantly boost our understanding of this class of Ly$\alpha$\ emitters and bring light to their still debated origin and formation. With a redshift of $\sim$3.3, the kinematics of the Ly$\alpha$ line will enhance our understanding of galaxies' gas inflows and outflows and the escape of Ly$\alpha$ photons at the start of the cosmic noon, the epoch of the peak of star-forming activity.
   \end{enumerate}

\begin{acknowledgements}
We thank the reviewer for his/her thorough review and highly appreciate the comments and suggestions, which significantly contributed to improving the quality of the publication. 
Based on observations obtained with MegaPrime/MegaCam, a joint project of CFHT and CEA/IRFU, at the Canada-France-Hawaii Telescope (CFHT) which is operated by the National Research Council (NRC) of Canada, the Institut National des Science de l'Univers of the Centre National de la Recherche Scientifique (CNRS) of France, and the University of Hawaii. This work is based in part on data products produced at Terapix available at the Canadian Astronomy Data Centre as part of the Canada-France-Hawaii Telescope Legacy Survey, a collaborative project of NRC and CNRS. Based on observations obtained with WIRCam, a joint project of CFHT, Taiwan, Korea, Canada, France, at the Canada-France-Hawaii Telescope (CFHT) which is operated by the National Research Council (NRC) of Canada, the Institut National des Sciences de l'Univers of the Centre National de la Recherche Scientifique of France, and the University of Hawaii.
This work is partly based on tools and data products produced by GAZPAR operated by CeSAM-LAM and IAP. The fits-file handling was done using the \cite{dfitspy} python module.
\end{acknowledgements}

\bibliographystyle{aa} 
\bibliography{bibliography} 


\end{document}